\documentclass[twocolumn]{aastex631}

\usepackage{amsmath}
\submitjournal{ApJ}

\shorttitle{Disk Flares due to Magnetic Reconnection}
\shortauthors{Fatuzzo et al.}
\begin{document}

\title{Avalanches and the Distribution of Reconnection Events in Magnetized  Circumstellar Disks }

\begin{abstract}
Cosmic rays produced by young stellar objects can potentially alter the ionization structure, heating budget, chemical composition, and accretion activity in circumstellar disks. The inner edges of these disks are truncated by strong magnetic fields, which can reconnect and produce flaring activity that accelerates cosmic radiation. The resulting cosmic rays can provide a source of ionization and produce spallation reactions that alter the composition of planetesimals. This reconnection and particle acceleration are analogous to the physical processes that produce flaring in and heating of  stellar coronae.  Flaring events on the surface of the Sun exhibit a power-law distribution of energy, reminiscent of those measured for Earthquakes and avalanches. Numerical lattice-reconnection models are capable of reproducing the observed power-law behavior of solar flares under the paradigm of self-organized criticality. One interpretation of these experiments is that the solar corona maintains a nonlinear attractor --- or ``critical''--- state by balancing  energy input via braided magnetic fields and  output via reconnection events. Motivated by these results, we generalize the lattice-reconnection formalism for applications in the truncation region of magnetized disks. Our numerical experiments demonstrate that these nonlinear dynamical systems are capable of both attaining and maintaining criticality in the presence of Keplerian shear and other complications. The resulting power-law spectrum of flare energies in the equilibrium attractor state is found to be nearly universal in magnetized disks. This finding indicates that magnetic reconnection and flaring in the inner regions of circumstellar disks occur in a manner similar to activity on stellar surfaces. These results, in turn, have ramifications for the spallation-driven injection of radionuclides in planetesimals, disk ionization, and the subsequent planetary formation process. 

\end{abstract} 
 
\author[0000-0002-1116-6140]{Marco Fatuzzo}
\affiliation{Physics Department, Xavier University, Cincinnati, OH 45207}

\author[0000-0002-8167-1767]{Fred~C.~Adams}
\affiliation{Physics Department, University of Michigan, Ann Arbor, MI 48109}
\affiliation{Astronomy Department, University of Michigan, Ann Arbor, MI 48109}

\author[0000-0002-9464-8101]{Adina~D.~Feinstein}
\altaffiliation{NSF Graduate Research Fellow}
\affiliation{Department of Astronomy and Astrophysics, University of Chicago, Chicago, IL 60637, USA}

\author[0000-0002-0726-6480]{Darryl Z. Seligman}
\affiliation{Department of Astronomy and Carl Sagan Institute, Cornell University, 122 Sciences Drive, Ithaca, NY, 14853, USA}


\section{Introduction} 
\label{sec:intro}

Circumstellar disks play a vital role in the formation of both stars and planets. Due to the large angular momentum of pre-collapse cores, the majority of star-forming material initially accretes onto the circumstellar disk \textit{before} accreting onto the star itself.  These processes define the initial conditions for the Keplerian environments where protoplanets and planets eventually form (e.g., see 
\citealt{Shu1987} to \citealt{mckeeost}).   

Observational and theoretical evidence indicates that disks are typically truncated at inner radii of order $R_x\sim10$ $R_\ast$. In these sufficiently ionized  regions   strong stellar poloidal magnetic fields that thread the inner disk  couple to the plasma motions \citep{Shu1994,Patterson1994,Akeson2005,DAlessio2005,Johns-Krull2009,Bate2009}.  In  regions with sufficient magnetic diffusivity, the Keplerian shear drives toroidal twisting of the initially poloidal fields, similar to the $\Omega$-effect in the Solar dynamo \citep{Parker1970}. The continuous winding and braiding of the poloidal fields leads to continuous reconnection events which produce the flaring activity \citep{Parker1972} that creates ionizing photons (XUV radiation) and powers the acceleration of cosmic rays (primarily protons). These cosmic rays, in turn, can drive nuclear spallation reactions \citep{Shu1996,Shu2001} that affect the composition and chemistry of planets forming in the region \citep{adams2021}. The goal of this paper is to explore models of flaring activity driven by reconnection events in the disk truncation region. Toward this end, we start with models of magnetic reconnection \citep{bak1987,bak1988,Lu1991,Morales2020} developed previously to explain the power-law distribution of flares on stellar surfaces \citep{Lu1993,Crosby1993,Aschwanden1998,Charbonneau2001,deArcangelis06} and adapt them for the magnetic truncation region of circumstellar disks. 

In this paper we focus on flaring events in the region of the disk near the magnetic truncation radius. Analogous flaring activity has been studied for the case of magnetic reconnection on stellar surfaces, including the Sun. As a working model of the process, the magnetic fields experiencing reconnection events on the solar surface have been proposed to reach a self-organized critical state \citep{bak1987}, where the magnetic field structures achieve a particular scale-invariant configuration \citep{Kadanoff1989,Babcock1990}. These types of models naturally explain the observed power-law distributions of the energy released during flares, the time duration of flare events, and the peak flux \citep{Lu1991}, and have been highly successful, both as an explanation for the observed stellar flaring activity (\citealt{Edney1998}; \citealt{Farhang_2018}; see also \citealt{Feinstein2022}) and as a classic example of self-organized criticality in action \citep{newmann2005}. Reconnection events, which occur at specific locations within the system, must propagate to nearby locations and instigate the release of magnetic energy at those sites. Under the right circumstances, this process of magnetic energy release leads to a local instability, with a cascade (or avalanche) of reconnection events that propagates through the system.  The entire  system can then be driven to a self-organized critical state that yields power-law distributions for the energy release, duration, and peak flux of the events. 

Our results indicate a universality in the power-law distributions obtained under different perturbation scenarios for the trigger scenario and redistribution rules adopted in our work.  It should be noted, however, that the reconnection criteria and redistribution rules used in avalanche models are ad hoc, and can produce different power-law spectral indices (e.g., \citealt{Edney1998, Farhang_2018}), although we note that the latter work invokes the principle of minimum energy so as to yield a definite model.  Furthermore, the adopted formalism is not a unique method to produce  a power-law distribution of energetic events \citep{Rosner1978,Litvinenko1996,Newman1996,Longcope1994,Einaudi1996,Galsgaard1996,Dmitruk1997,Galtier1998,Georgoulis1998,Einaudi1999,Galtier1999}. 

The importance of possible flaring activity at the inner truncation edge of the disk warrants an examination of the underlying reconnection processes in disk environments, which provide somewhat different baseline conditions than stellar surfaces. More specifically, the magnetic fields near the inner disk edge are strongly affected by differential rotation of the disk, i.e., by Keplerian shear. On one hand, the shearing motions provide a mechanism for twisting up magnetic field lines, and thereby instigate the magnetic reconnection events. On the other hand, the shear could in principle spread out the field lines and thereby inhibit reconnection. 

The magnetic truncation region, which defines the inner boundary of the circumstellar disk, is of interest for several reasons.  Accretion from the disk onto the star occurs through this truncation region, so that the any flaring activity can affect the ionization state and chemical properties of the accreted material \citep{Shu1994}.  The rocky material that is left behind -- that does not accrete onto the star -- can be enriched via spallation reactions \citep{adams2021}. The subsequent enirchment with radioactive nuclei  can affect planetesimal properties \citep{Urey1955,Reiter2020} and may lead to melting \citep{Schramm1971,Hevey2006}, differentiation \citep{LaTourrette1998,Moskovitz2011} and volatile removal or dehydration \citep{Grimm1993,Ikoma2018,Lichtenberg2019}. A large fraction of the planetary population observed in transit resides in this general region, including the subset of planets that display surprising regular properties, sometime called peas-in-a-pod patterns \citep{weiss2018,weiss2022}. Finally, we note that Hot Jupiters also reside near the location of the disk truncation boundary (e.g., see \citealt{dawson2018}), so that flaring activity can in principle affect their composition and ionization properties. 

We note that spallation reactions driven by cosmic rays --- one motivation of this study ---  have been invoked previously in a number of contexts. As outlined above, spallation from cosmic rays produced in the disk truncation region could account for the anomalous abundances of short-lived nuclei inferred for the early Solar Nebula \citep{Shu1996,Gounelle2001,Leya2003,Duprat2007}, including $^{26}$Al and $^{10}$Be. In a more general context, such radioactive enrichment can affect planets forming near the truncation region in other planetary systems by providing an additional source of both heating and ionization \citep{adams2021}. Cosmic rays can also be accelerated by protostellar shocks \citep{Hayashi1965}, and will subsequently contribute to both ionization and nuclear spallation reactions \citep{padovani2016,gaches2020}. For completeness, we note that there is no current consensus on whether radionuclides were injected into the protosolar nebula via local spallation reactions or by external sources (or both). More specifically, an external stellar source for the inferred short-lived radionuclides -- such as supernovae in the solar birth cluster -- has been widely considered (e.g., 
\citealt{Goswami2005,Krot2009,Cameron1977,Hester2004}; see also \citealt{adams2010} for a review). For any source of cosmic rays --- either external or produced locally via shocks or reconnection events --- the shearing of magnetic fields within the disk affects their propagation and retention \citep{fujii2022}.

Given that flaring activity near the inner truncation edge of the disk can potentially affect planet formation, the goal of this paper is to study how magnetic reconnection in the disk environment differs from that on the stellar surface, with particular attention focused on whether the presence of a Keplerian shear affects the resulting distributions of flare energies and durations. As outlined above, we use the well-established model of \cite{Lu1991} as a starting point. In Section 2, we outline the scales of the problem and the properties of the disk environment under consideration. In Section 3, we review the previous model (see also \citealt{Lu1993}) and adapt it to the conditions expected in the truncation region of circumstellar disks. The properties of the self-critical state are then outlined in Section 4. With this formulation in place, we find the distributions of flare energies, duration, and power for a range of parameter space in Section 5. The paper concludes, in Section 6, with a summary of our results and a discussion of their implications.  
 

\begin{figure*}[]
\begin{center}
       \includegraphics[scale=0.5,angle=0]{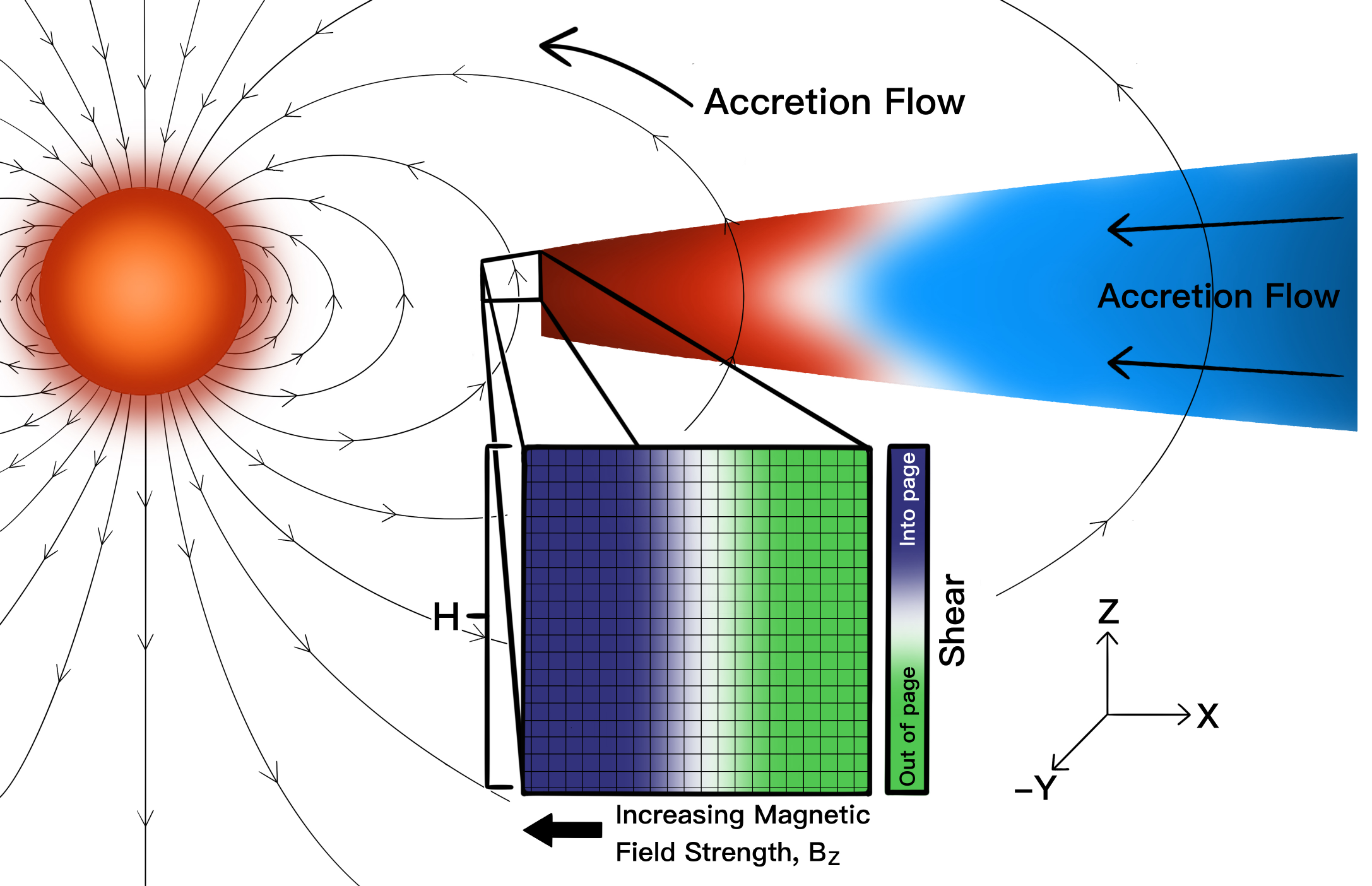}
    \caption{A schematic diagram of the physical situation investigated in this paper.}\label{fig:geometry} 
\end{center}
\end{figure*}

\section{Scales} \label{sec:scales}

\subsection{Length Scales}
The magnetic truncation region is the source of magnetic reconnection events, which lead to the production of intense cosmic radiation. This inner boundary of the disk is determined by the balance between the inward ram pressure of the accretion flow and the outward magnetic pressure from stellar magnetic field lines. Young stellar objects typically have surface fields 
$B_*\sim1-3$ kG with disk accretion rates ${\dot M}\sim10^{-8}M_\odot$/yr \citep{Johns2007}, which lead to truncation radii in the range $R_x\sim7-10R_\ast$. 

The other relevant length scale in this problem is the scale height $H$ of the disk, where $H/r\sim1/20$. The reconnection events of interest must take place on a spatial scale $\ell$ that is small compared to the disk scale height. We thus have the ordering of length scales given by 
\begin{equation}\label{eq:length_ordering}
    \ell \ll H \sim R_\ast \ll R_x\,. 
\end{equation}
With this ordering, we assume that the forthcoming simulation volume is a cubic lattice with side length equal to the disk scale height $H$, comprised of $N^3$ cubic cells with side length $\ell = H/N$ within which the reconnection events occur.  If we take the active region where magnetic reconnection occurs as a toroid of inner radius $R_x/2$, outer radius $R_x$, and total height $2H$, then the number of lattice cubes in which flaring events can be triggered is found to be 
\begin{equation}
    N_L\equiv {3\pi\over 2} {R_x^2 \over H^2}=600\pi \left({R_x\over 20H}\right)^2\,.
\end{equation}

We note that on the lattice scale, the curvature of the magnetic field lines is small, ${\cal O}(H/r)$, so that the background magnetic fields, which determine the outer boundary condition, can be taken to be vertical (see also the following section). Similarly, the curvature in the azimuthal direction is also of order ${\cal O}(H/r)$.

A schematic representation of this configuration is shown in Figure \ref{fig:geometry}. Within the resulting cubic grid, the most important generalization from previous work is the inclusion of Keplerian shear in the background velocity, along with a corresponding gradient in the background magnetic field strength (where both vary in the radial or ${\hat x}$ direction). The largest spatial scale that is resolved in this treatment is the scale height of the disk. The cascade of flares can be resolved down to a much smaller scale set by the choice of grid size $\ell$, as discussed in the following section. 

\subsection{Timescales}

Next we consider the relevant time scales. With typical truncation radii $R_x\sim0.07$ AU, the orbital time scale $P_{orb}$ is a few days (comparable to the orbital periods of Hot Jupiters). The sound crossing time of the simulation volume may be calculated $t_S=H/c_S$, where $c_S\sim 4$ km/s is the thermal sound speed at the expected temperatures of $T\sim2000$ K. The sound crossing time scale is therefore roughly comparable to the orbital time scale.

In contrast, the timescale over which magnetic reconnection occurs is given by the Alfv\'en crossing time over the turbulance lengthscale, which in our case is given by the grid length $\ell$.  As a benchmark, the unamplified magnetic field strength near the disk edge is expected to be of order $B\sim3$ G (assuming a stellar surface field of $\sim3$ kG and a dipole radial dependence). The density of the reconnection region is expected to be given by
\begin{equation}
    \rho\simeq 10^{-15}\,{\rm g/cm^3} \,\bigg(\,\frac{\Sigma}{10^{-4}\;{\rm g/cm^2}} \, \bigg) \, \bigg( \,\frac{H}{5\times 10^{10}\,{\rm cm}}\,\bigg)^{-1}\,.
\end{equation}
The benchmark value of the column density is inferred from observations, which show that X-ray emission (assumed here to be coincident with the reconnection region) in young stellar objects takes place during optically thin conditions \citep{Lee1998}. As a result, the Alfv\'en speed has the form 
\begin{equation}
    v_A = {B\over \sqrt{4\pi\rho}} \sim 300 \,{\rm km/s} \left({B\over3\,{\rm G}}\right)\,\bigg(\frac{10^{-15}\,{\rm g/cm^3} }{\rho}\,\bigg)^{-1/2} \,.
\end{equation}
Even if the magnetic field in the disk were simply the stellar field, the Alfv\'en speed $v_A$ is considerably larger than the sound speed,
resulting in a reconnection time much shorter than $t_S$. In addition, however, the magnetic field within the disk will be amplified by the rotational winding induced dynamo.  An assessment of the magnetic field strengths required to produce the observed luminosities from young stellar systems is presented in Section 5.1, where the obtained estimate of $B\sim100$ G translates to $v_A\sim10^4$ km/s. As a result, the ordering of velocities in the problem is given by 
\begin{equation}
    v_A  \gg c_S \, .
\end{equation}

\subsection{Ordering of Injection and Reconnection Timescales}

A useful parameter to characterize the relative importance of Alfv\'enic and acoustic perturbations is the plasma $\beta$ parameter, defined according to 
\begin{equation}
    \beta \equiv 2\frac{c_S^2}{v_A^2}\,.
\end{equation}
 When $\beta>1$ acoustic modes dominate both the dynamics and energy transport, and when $\beta<1$ Alfv\'enic modes dominate.  In the case of the Solar corona, $\beta<1$ in the outer atmosphere and $\beta\sim1$ in the inner atmosphere. As a result, convection and/or differential rotation couple to  Alfv\'enic modes in the corona and drive heating \citep{Ionson1985}. 
 
 Relevant timescales are the timescales of subsurface convection injecting braided fields into the corona and the Alfv\'enic crossing timescale. When these timescales are well seperated, braided magnetic fields build up and can trigger avalanching events. This is known as the direct current or DC coronal heating regime. 

 The analogy holds in the truncation region in the disk. Here, the injection timescale is expected to be comparable to the oribital period, and the reconnection timescales are observed to be on the order of minutes. The buildup of magnetic perturbations can therefore occur incrementally before an avalanche of reconnection events occur.  


\section{Methodology} \label{sec:method}

We model the avalanche of magnetic reconnection events in the inner disk region of young stellar objects using a lattice model of magnetic energy release in solar flares developed previously (by \citealt{Lu1991}; see also \citealt{Lu1993, Edney1998}).  In order to simulate the effects of a Keplerian shear on the reconnection environment, we expand on these previous treatments by considering various sheared magnetic perturbation and reconnection scenarios as described below. 

Motivated by the discussion of length scales and timescales in Section \ref{sec:scales}, we consider a representative cubic lattice of side length $H$ located within the flaring region of the system, as shown in Figure \ref{fig:geometry}.  The lattice, in turn, is comprised of $N^3$ cells of side length $\ell=H/N$.  
For the adopted scaling $H = R_x/20$, the stellar magnetic field threading a cubic lattice whose inner boundary (which sets $x = 0$) is a radial distance $R_{in}\gg R_*$ from the stellar center is well approximated by the expression
\begin{equation}
    {\bf B_0} =  B_* \left({R_*\over R_{in}}\right)^3\; \left(1-{3\over 20} {x\over H}\right)\hat z\,.\label{eq:starfield} 
\end{equation}
Here we are assuming that the dipole term dominates the stellar field at the location of the simulation volume. 

A self-organized critical state is maintained in the lattice by two types of events: (i) the injection of asymmetric magnetic perturbations $\delta \bf B$ that drive the system, and (ii) the triggering of reconnection events that release magnetic energy, and thereby relax the system. 
In our model, perturbations are added step-wise to a randomly selected cell in the lattice, where each ``step" occurs over a time interval $\Delta t_p$. The three Cartesian components of $\bf \delta B$  are randomized using flat-top distributions of width $\Delta b$, with the $\hat y$ component off-centered by an amount $b_0$ so as to model asymmetric magnetic fluctuations driven by disk rotation (see also \citealt{Lu1991,Lu1993} for additional detail). 

Magnetic reconnection occurs when the local magnetic field curvature exceeds a critical threshold $B_c$ as defined by the condition
\begin{equation}
|d{\bf B_{{\bf n}}}|=\Big|{\bf B_n}-{1\over 6}\sum_{{\bf m}} {\bf B_{n+m}}\Big| > B_c\,,\label{eq:discretepoisson} 
\end{equation}
where ${\bf n} = \{n_x, n_y, n_z\}$ denotes a specific cell in accordance to a standard labeling scheme (i.e., the $n$ indices run from 1 to $N$), and $\bf m$ runs over the six unit vectors $\{\pm 1,0,0\}$, $\{0,\pm 1,0\}$, and $\{0,0, \pm 1\}$ that represent the nearest neighbors.  (Here we are using the subscripts to denote the vector location following  \citealt{Lu1991,Lu1993}.) The magnetic field in a triggered cell and its nearest neighbors is readjusted via the reconnection rules
\begin{equation}
    {\bf B_n} \rightarrow {\bf B_n} -{6\over 7} \; {d{\bf B_n}\over |d{\bf B_n}|}B_c \, ,\label{eq:rule1} 
\end{equation}
and
\begin{equation}
     {\bf B_{n+m}} \rightarrow {\bf B_{n+m}}+{1\over 7} \; {d{\bf B_n}\over |d{\bf B_n}|}B_c \,.\label{eq:rule2} 
\end{equation}

We note that the addition of a single perturbation to a cell in the lattice violates the divergence-free condition ${\bf \nabla \cdot \delta B}=0$, indicating that this class of models is too idealized to accurately represent the complex structure of a turbulent magnetic field at the granular scale of a cell.  However, the mean magnetic field added to the lattice is uniform, and therefore divergenceless over the lattice length scale and the flaring time scale. Note that, globally, the mean magnetic field added to the entire system is azimuthal and constant spatially, and is therefore also divergence-free. In contrast, the readjustment of the magnetic field, which simply redistributes part of the magnetic field of a "central" cell evenly to its six neighbors, is divergenceless since no field was added or removed for a region containing all seven cells.  Notice, however, that field readjustments at fixed (Dirichlet) boundaries do lead to the removal of magnetic field from the lattice.

Although a readjustment conserves the magnetic field (with the exception of boundary cells), the nonlinearity of magnetic energy ($U\propto B^2$) coupled with the curvature condition for reconnection to occur results in a loss of magnetic energy in the triggered cell that exceeds the net gain within the six neighboring cells, and leads to an overall net energy loss for each triggered event given by 
\begin{equation}
    E_0 = {6\over 7}\,{[B_c]^2\over 8 \pi}\left(2 {|d{\bf B_n}|\over B^2_c}-1\right)\ell^3 \,.\label{eq:energy} 
\end{equation}
In the $\delta B\ll B_c$ limit used in all simulations presented in this work, the magnetic field curvature for a triggered cell just exceeds the critical threshold, so that the energy lost for every trigger event is well approximated by the expression 
\begin{equation}
E_0 \simeq {3\over 28}{B_c^2\over \pi} \ell^3\,.
\end{equation}

Since the system resides in a self-critical state, a readjustment in the field can trigger further magnetic reconnection events. This process can manifest as an avalanche-like effect that propagates neighbor-to-neighbor throughout the lattice  (including reflections at the boundaries). As a result, the triggering condition given by Equation (\ref{eq:discretepoisson}) is evaluated after every readjustment within each lattice cell for which the magnetic field changed. If the triggering condition is not met, then time is advanced by $\Delta  t_p$, and the process of adding perturbations to the lattice and checking for trigger events is repeated.  When reconnection events are triggered, the magnetic field throughout the lattice is simultaneously readjusted according to equations (\ref{eq:rule1}) and (\ref{eq:rule2}), and the avalanche time is advanced by an interval $\Delta t_a$ corresponding to the time it takes for the neighboring cells to be affected. All affected cells are then checked to see if they meet the trigger threshold, and if any are triggered, the avalanche is propagated until the system is fully relaxed.  The  avalanche event is therefore completed  when no cells satisfy the triggering criteria after a readjustment.   


The avalanche problem contains two relevant timescales --- (i) the perturbation injection time-step $\Delta t_p$ and (ii) the avalanche propagation time-step $\Delta t_a$.  The model described above assumes a ``fast" reconnection scenario in which avalanche events occurs entirely within a perturbation time-step $\Delta t_p$. As shown in Section 5, such a scenario is not completely consistent with observations for the longest lasting avalanches, i.e., for the tail of the distribution. As such, we will also consider a ``slow" reconnection scenario in which $\Delta t_p = \Delta t_a$ (so that a perturbation is added randomly to the lattice at each avalanche step) to gain insight into how the addition of magnetic perturbations during an avalanche event may affect the resulting output measures.

Regardless of which reconnection scenario is chosen (fast or slow), the procedure described above drives any initial magnetic configuration of the lattice to its self-critical state.  We evaluate three output measures that characterize the avalanche events once the system reaches equilibrium. The first is the avalanche time $T_A = N_a \Delta t_a$, where $N_a$ represents the total number of avalanche steps required for the triggered system to reach a relaxed state.  The second is the total magnetic energy $E_T \equiv N_t E_0$ lost during an avalanche, where $N_t$ is the number of times the trigger threshold was exceeded within a given avalanche (and in turn, the number of times the magnetic field was redistributed).  We note that the readjustment process makes it possible for a given cell to be triggered multiple times during an avalanche, and as such, $N_t$ is not limited by the number of cells in the lattice. The third measure is the maximum power $P_M$, defined as the maximum of the set of power output values $P\equiv N_p E_0 /\Delta t_a$ generated during the avalanche event, where $N_p$ is the number of cells that are triggered during each respective time step $\Delta t_a$ of that event. The frequency distributions of these output measures are used to characterize the flaring process on a global scale. 

For completeness, one can also consider the waiting time distribution, essentially the distribution of time intervals between events. In general, models of self-organized criticality tend to produce exponential waiting time distributions. If a systems is driven by a non-stationary random process, however, the resulting frequency distribution of waiting times often includes a power-law tail \citep{norman2001}. We have calculated the waiting time distribution for our model and find that it has an exponential form (as expected) with some indication of a power-law tail, although more modeling is necessary to fully explore this regime. On the observational front, the flaring activity from the Sun shows a power-law tail, but the waiting time distributions for flares from other stars have not been measured.  

\section{The Self-Critical State}

The random injection of magnetic perturbations within the lattice leads to a self-critical state comprised of a fluctuating field superimposed onto a static field. The structure of this imposed field is governed by the trigger criteria and the magnetic field boundary conditions of the lattice.  The results of our simulations indicate that any initial (non-critical) magnetic profile adopted for the lattice will be driven to that self-critical state, regardless of the structure of the magnetic perturbations (as defined by $b_0$ and $\Delta b$ in our model; see Section 3) that drive it.  We remind the reader that a self-critical state is only reached if non-asymmetric perturbations are added.  As a result, the only component of the static field that has a nontrivial structure is the one to which $b_0$ is included (which for our scenario is the azimuthal or $y$ component).  

The expression for the local magnetic field curvature used in equation (\ref{eq:discretepoisson}) is  the finite-difference form of the Laplacian operator applied to the magnetic field (but divided by a factor of 6):
\begin{equation}
   \;d{\bf B_n}=  -{\ell^2\over 6}\; \nabla^2{\bf B} 
   \Big|_{fd}\,,
\end{equation}
where the subscript denotes the finite-difference version of the operator. As a result, it is not surprising that our simulations show that the underlying static structure of the field can be obtained from the solution to the Poisson equation 
\begin{equation}
    \nabla^2 B = -\eta \,{6 B_c\over \ell^2}\,,\label{eq:poisson} 
\end{equation}
with appropriate boundary conditions imposed.  The value of $\eta=1/\sqrt{3}$ manifests for all cases considered in this work. The field $B$ corresponds to the $y-$component of the magnetic field (which is the component with nontrivial structure).  

The ability to find the smooth surface that characterizes a self-critical state provides a computationally expedient way of generating an initial magnetic profile for the lattice that is efficiently driven to a self-critical state by the addition of magnetic perturbations.  However, we stress that the highly idealized nature of our model does obscure the physical interpretation of these smooth structures.  In particular, we find that the maximum strength of the smooth magnetic field for our simulations is of order $B_{max}\simeq 600 \,B_c$, and the mean value for the smooth field over the entire lattice is $\langle B\rangle\simeq 330 \,B_c$.  In contrast, the perturbations added to each cell have magnitudes $\delta B\ll B_c$.  As a result, the mean magnetic energy added to a cell by the addition of a magnetic perturbation (which scales as $\langle B\rangle\cdot \delta B$) is less than the magnetic energy stored in the cell (which scales as $\langle B^2\rangle$) by a factor $\simeq 300$. As shown in Section 5.1, however, in order for the energy in the perturbed magnetic field to be large enough to account for the expected reconnection luminosity (inferred from X-ray observations), the background field cannot be as large as that suggested by the surfaces shown in Figure 2.  
In short, these surfaces represent the shape of the "sandpile" needed to attain the self-critical state, which nonetheless provide a useful heuristic picture of the self-critical state.  
The physical magnetic field is thus represented by a fluctuating field superimposed on these surfaces.

\section{Results}
\subsection{Fast Reconnection model}

Following prior works (again see \citealt{Lu1991} and related references), we first consider a scenario were avalanche events occur within the injection time-step $\Delta t_p$. This  scenario is defined by $N = 50$, a dimensionless trigger threshold  $B_c = 1$, and dimensionless perturbation parameters $\Delta b = 0.13 B_c$ and $b_0 = 0.035 B_c$.  As noted above, the offset $b_0$ is only included for the $y-$component of $\delta {\bf B}$, as may be expected for magnetic fluctuations that are driven by the rotation of the disk.  
Because of this only the $y$-component of the magnetic field will reach a self-critical state.  Motivated by the disk geometry, we adopt periodic boundary conditions in the azimuthal ($y$) direction, and set the magnetic field to zero at the other four boundaries. (The stellar magnetic field that threads the lattice is not included in this initial case, but will be considered in a following simulation.)  

\begin{figure*}[]
\begin{center}
       \includegraphics[scale=0.7,angle=0,trim={5cm 7cm 5cm 7cm}]{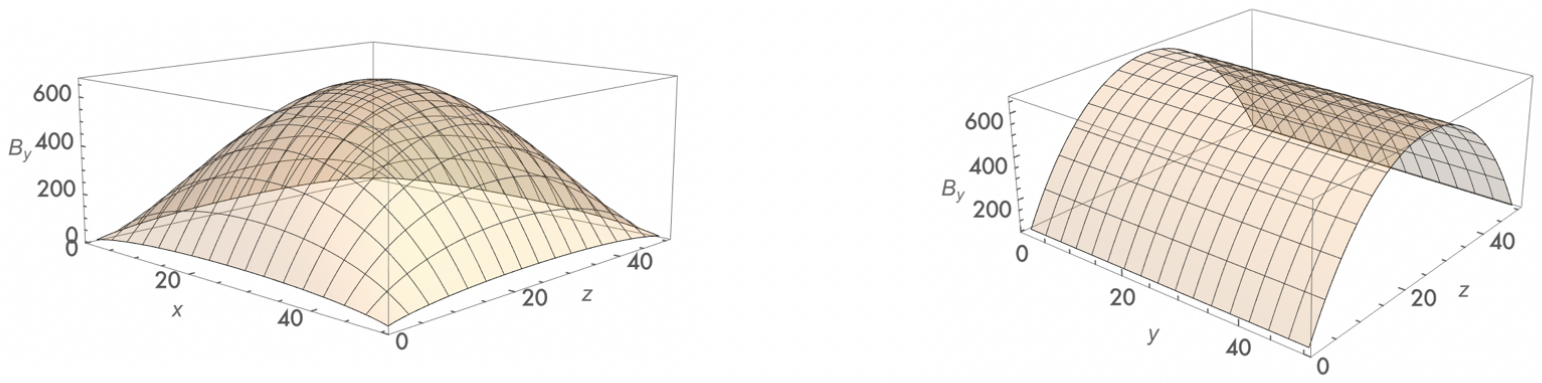}
    \caption{Smooth component of the critical magnetic field $B_y$ for our baseline model.  The left panel shows a cut along the $x-z$ plane, and the right panel shows a cut along the $y-z$ plane.}\label{fig:surface1} 
\end{center}
\end{figure*}

\begin{figure}[]
\begin{center}
       \includegraphics[scale=0.35,angle=0]{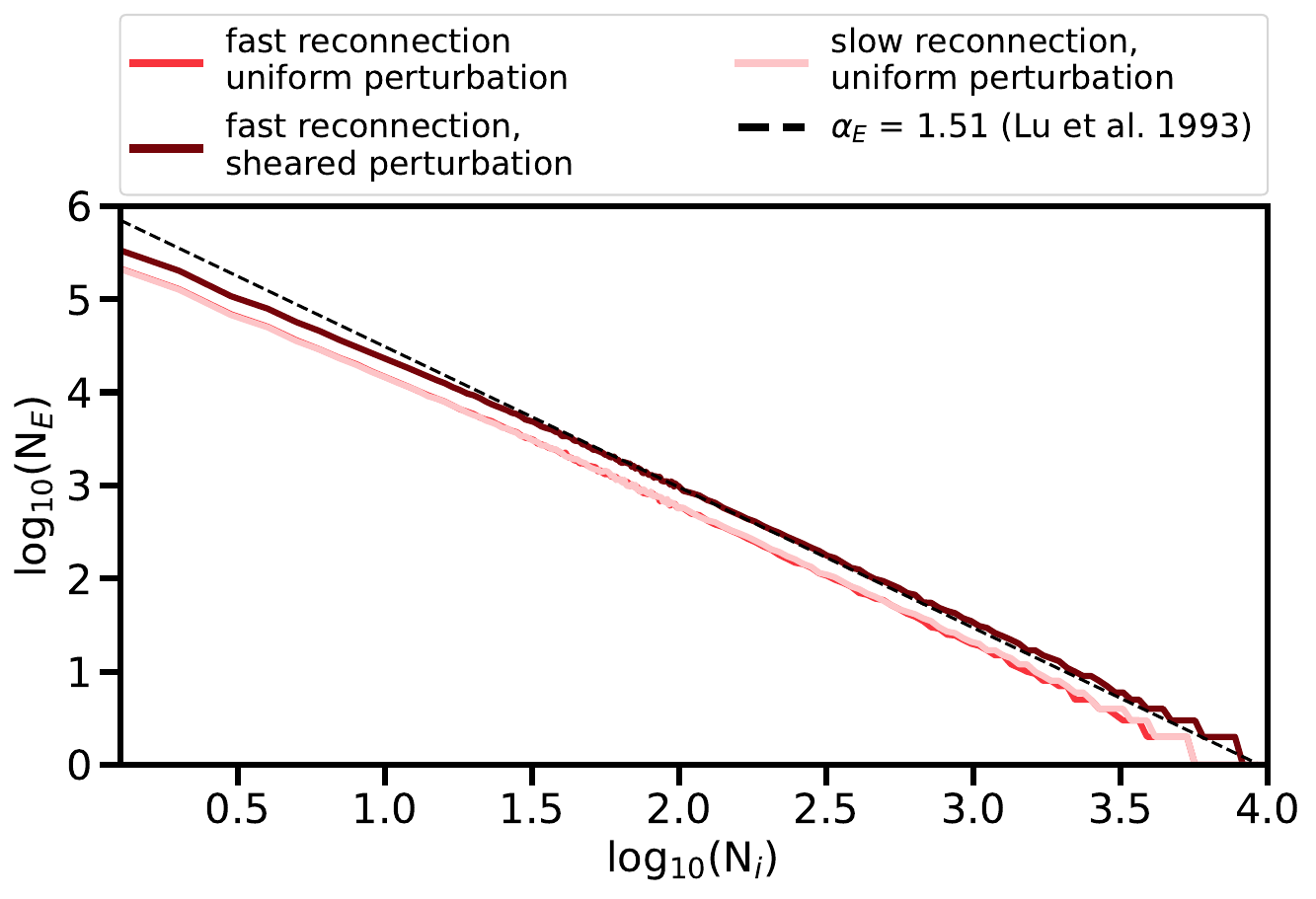}
       \includegraphics[scale=0.35,angle=0]{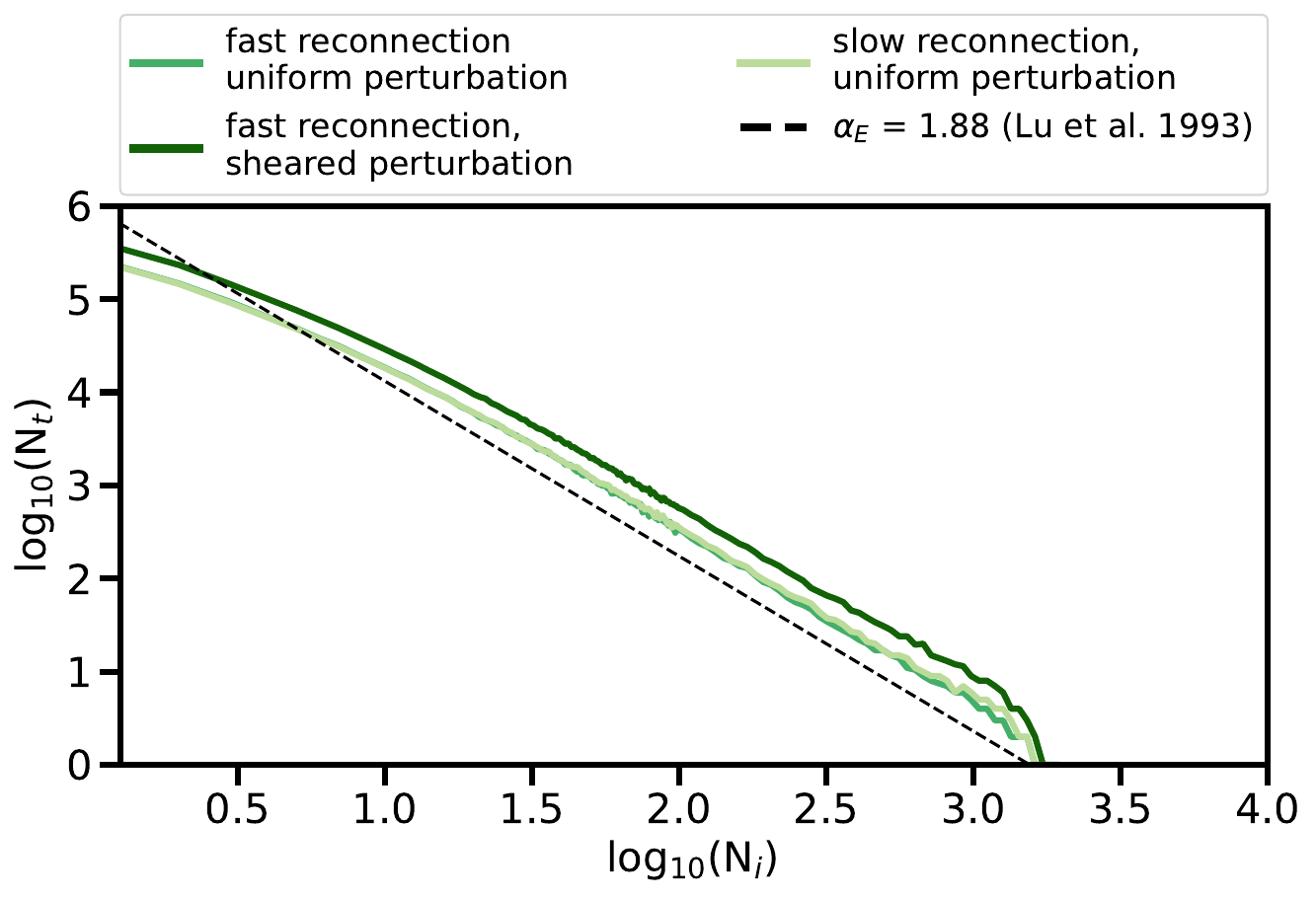}
       \includegraphics[scale=0.35,angle=0]{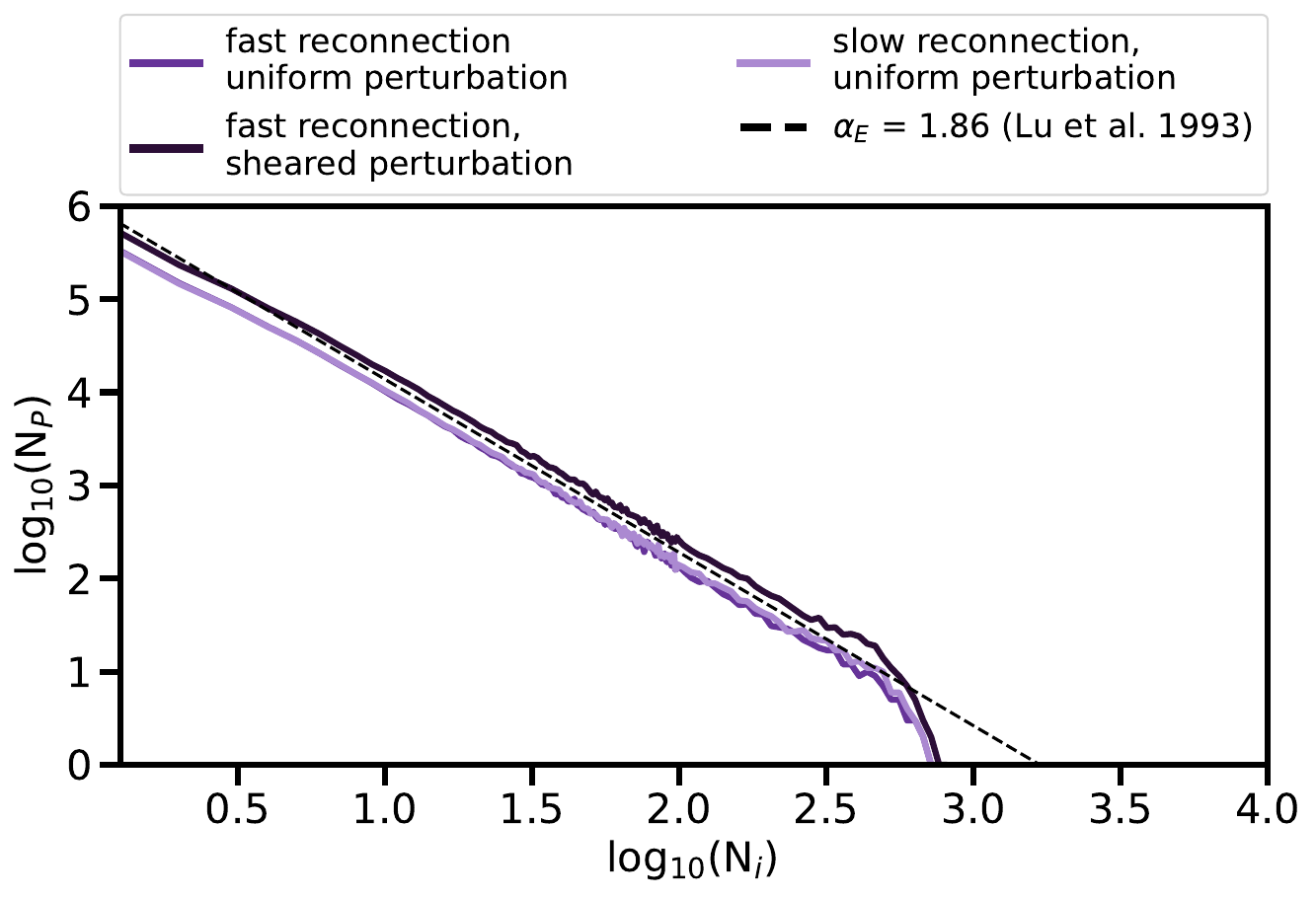}
    \caption{Frequency distributions of avalanche total energy $N_E$ for the fast reconnection model (red curve), slow reconnection model with uniform perturbation structure (maroon curve), slow reconnection model with a sheared perturbation structure (pink curve). We show the corresponding distributions for duration $N_t$ and power $N_P$ in green (middle) and purple (bottom) as well. In each panel, the dashed black line represents the corresponding power-law fit from \cite{Lu1993}.}\label{fig:comparen} 
\end{center}
\end{figure}

The structure of the smooth magnetic field component for this state obtained from equation (\ref{eq:poisson}) is illustrated in Figure \ref{fig:surface1}.  Once a self-critical state was reached in our simulation, a total of $10^8$ perturbations were added to the lattice, resulting in a total of $1.6\times 10^6$ avalanche events. The resulting mean time between avalanche events was $\simeq 60 \Delta t_p$.  The frequency distributions for our three output measures are presented in Figure \ref{fig:comparen}, along with the power-law fits from \cite{Lu1993}.  While our results are in good agreement with those of \cite{Lu1993}, we note that the frequency distribution of our energy output measures is in slightly better agreement with the results of \cite{Edney1998}, for which $\alpha_E=1.45\pm0.04$.

Of importance in our upcoming analysis is the expectation value $\langle N_t\rangle$ of the energy frequency distribution.  Since the largest avalanches produce most of the energy, the value of $\langle N_t\rangle$ depends sensitively on where the frequency distribution $E_T$ truncates. These numerical experiments  do not have a sufficient number of  events to resolve the turnover at the low end of the energy spectrum.  However, an upper limit can be established by noting that the turnovers for the duration and maximum power are resolved. These occur at $N_{t;max}\simeq 1600$ and $N_{P;max}\simeq 600$, respectively, which in turn establish $N_{E;max}= N_{t;max}\cdot N_{P;max}\simeq 10^6$.  The power-law fits to our data can then be used to obtain a value of
\begin{equation}
    \langle N_t\rangle = {\int_1^{10^6}N^{-0.51}dN\over {\int_1^{10^6}N^{-1.51}dN}}\approx 900\,.
\end{equation}

Following \citet{Lu1993}, we assume that the energy released in a reconnection event is given by 
\begin{equation}
    \Delta E=\ell^3{\langle B_\perp^2\rangle\over 8\pi}\,,
\end{equation}
where $B_\perp\simeq B/2$, and that each elementary reconnection occurs during a time
\begin{equation}
    \Delta T={\ell\over v_A}\xi\,,
\end{equation}
where $\xi\sim 10 - 100$ \citep{Parker1979}.

We assume a field with strength $B\simeq 100$ G, which is consistent with solar flare values. This  would result from a modest amplification of the background stellar field (with an expected average $B \sim $ 10 G) in the active region by the rotational motion of the plasma. With this benchmark value, the change in energy is estimated as $\Delta E \simeq 10^{29}$ ergs over a  time scale $\Delta T \simeq 10$ s for $\xi = 10$.  The mean energy released during the avalanche events presented in Figure 3 for the fast reconnection model is 
$\langle E_a\rangle=\langle N_t\rangle\Delta E\approx 9\times 10^{31}$ ergs.
If we further assume that each lattice undergoes an avalanche event on a timescale comparable to the orbital period, which we take to be 4 days, then the overall luminosity generated within the entire magnetic reconnection region is
\begin{equation}
    L={N_L\langle E_a\rangle\over P_{orb}}\simeq 5\times 10^{29}\,{\rm erg/s}\,,
\end{equation}
for the values of $N_L$, $\langle E_a\rangle$ and $P_{orb}$ used above.
This result is in good agreement with expected cosmic ray luminosities. Briefly, in young stellar objects, we expect the cosmic ray luminosity to be a significant fraction of the X-ray luminosity from flares, where the latter luminosity is typically 1000 times smaller than the photon luminosity \citep{feigelson2005,preibisch2005,padovani2016}, although the observations show a wide variation (both in time and from source to source). 

The results are also consistent with a global energy analysis. Specifically, the magnetic energy stored in the active region of the reconnection annulus is given by the integral of the energy density over the relevant volume
\begin{equation}
    E_B = 2 H \int_{R_x/2}^{R_x} {B_x^2\over 8\pi} \left({R_x\over r}\right)^6 2\pi r dr  = {3 B_x^2 R_x^3\over 32}\,,
\end{equation}
where we have used $H = R_x/20$. With $R_x = 10^{12}$ cm and $B_x = 3\,\rm G$,  the magnetic energy stored in the field is
$E_B = 8.4 \times 10^{35}$ ergs. This value is roughly $5$ times greater than the energy released during an orbital period.  The stellar dynamo coupled to disk rotation must therefore replenish $\simeq$ 20\% of the magnetic  energy during one  orbital period (at the location of the inner disk edge $r \sim R_x$). 

As noted above, an avalanche occurred on average every 60 perturbation steps in our simulation. Therefore, a perturbation is added to the lattice every $\Delta t_p\sim P_{orb}/(60)\simeq 5800$ s.  In contrast,
$\Delta t_a\sim \Delta T$, so that avalanches 
with $N_a \le 580$ steps would be expected to be completed before the next perturbation is added to the lattice.  The middle panel in Figure \ref{fig:comparen} shows that the vast majority of avalanche events satisfy this condition.  However, the longest and most energetic events will have a small number of perturbations added to them before the system is completely relaxed.

To conclude this section, we note that in order for the cosmic ray luminosity to be of order $\sim10^{29}$ erg/s (as inferred from observations -- see Section 5.1), the expected relationship that the perpendicular magnetic field $B_\perp$ (that being removed in a reconnection event) is roughly half of the total field $B$.  In contrast, the magnetic field in the simulation that is redistributed is of order $B_c$, which is much smaller than the total magnetic field found in the cell.  This result reaffirms our discussion in Section 4 noting that the smooth field that characterizes the self-critical state -- namely the field represented by the surfaces shown in Figure 2 -- provides only a heuristic description. The actual fields that one finds in flaring environments correspond to fluctuations superimposed on the aforementioned surfaces. Of course, the true configuration of the magnetic fields in these systems has not yet been measured, and such future observations will provide valuable information regarding their structure. 

\subsection{Sheared models}
In this subsection, we investigate the effect of augmenting the previous simulations with a background Keplearian shear. We  consider three distinct scenarios for which perturbations are injected into the lattice with an underlying sheared structure. These scenarios are motivated by both the structure of the stellar magnetic field that threads the (disk) active region and the differential Keplerian rotation of the disk. In the first gradient-in-perturbation scenario, magnetic perturbations are added that reflect the magnitude of the stellar magnetic field threading the lattice. Specifically, we inject the strongest perturbations on the  side of the lattice closest to the star. The magnitude of the perturbations decrease in accordance with the diminishing magnetic field at larger radii.   We exaggerate this effect by multiplying both $\Delta b$ and $b_0$ by a scaling factor g under the following rules:
\begin{equation}
\begin{cases}
\Delta b \rightarrow g \, \Delta b \\
b_0 \rightarrow g \, b_0\\
   g= \bigg(1-0.5 \left[{n_x-0.5\over N}\right]\bigg)\,.\\

   \end{cases}
\end{equation}
A non-dimensional magnetic field 
\begin{equation}
    \vec B_d=\left(1-0.15 \left[{n_x-0.5\over N}\right]\right)\hat z\,,
\end{equation}
representing the stellar magnetic field is maintained throughout the lattice (including the boundaries) for all three scenarios considered in this section.

 The scaling factor does not modify the smooth component of the critical magnetic field because the triggering field is not changed. Our results, depicted by the heavy solid curves in Figure \ref{fig:comparen}, indicate that the slopes of the output measure distributions are also not affected by the sheared perturbation structure. However, due to the reduction in the perturbation amplitudes, the average time between avalanche events increased from $60\Delta t_p$ to $100\Delta t_p$.

Another possible complication is that the perturbations to the magnetic field could have non-trivial structure, such as a well-defined (and non-zero) curl, which would result from an underlying current. To consider this second current-injection scenario, two perturbations were added within a time-step $\Delta t_p$, and correlated to produce a curled-field configuration. Specifically, the first perturbation was chosen as in  Section 5.1.  The second, correlated perturbation was  placed in the adjacent cell $(n_x,n_y,n_z)\rightarrow (n_x+1,n_y,n_z)$ with a  $y-$component obeying the rule 
\begin{equation}
    \delta B_{y;n_x+1}=\begin{cases}
-\delta B_{y;n_x}/2\qquad \delta B_{y;n_x}>0\\
   2 \delta  B_{y;n_x}\qquad \delta B_{y;n_x}<0\,.\\

   \end{cases}
\end{equation}
As before, 
our results (not shown) indicate that the slopes of the output measure distributions are not affected by the presence of a curl in the injected magnetic field perturbation.

For the third sheared-trigger scenario, we adopt a critical field value that reflects the magnitude of the stellar magnetic field threading the lattice, but again exaggerate the effect by using the form
\begin{equation}
\begin{cases}
B_c \rightarrow g \, B_c \\
   g= \bigg(1-0.5 \left[{n_x-0.5\over N}\right]\bigg)\,.\\

   \end{cases}
\end{equation}

We note that doing so affects both the perturbation parameters $\Delta b$ and $b_0$ as well as the magnetic energy released during the redistribution of the magnetic field. The spatially dependent trigger threshold also affects the smooth structure of the self-critical fields, skewing it somewhat toward lower values of $x$. As with the other two cases considered in this section, our results (not shown) indicated that there is no significant effect on the slopes of output distributions from a sheared trigger threshhold.

\begin{figure*}[!bt]
\begin{center}
       \includegraphics[scale=0.5, trim={0cm 0cm 0cm 1cm}]{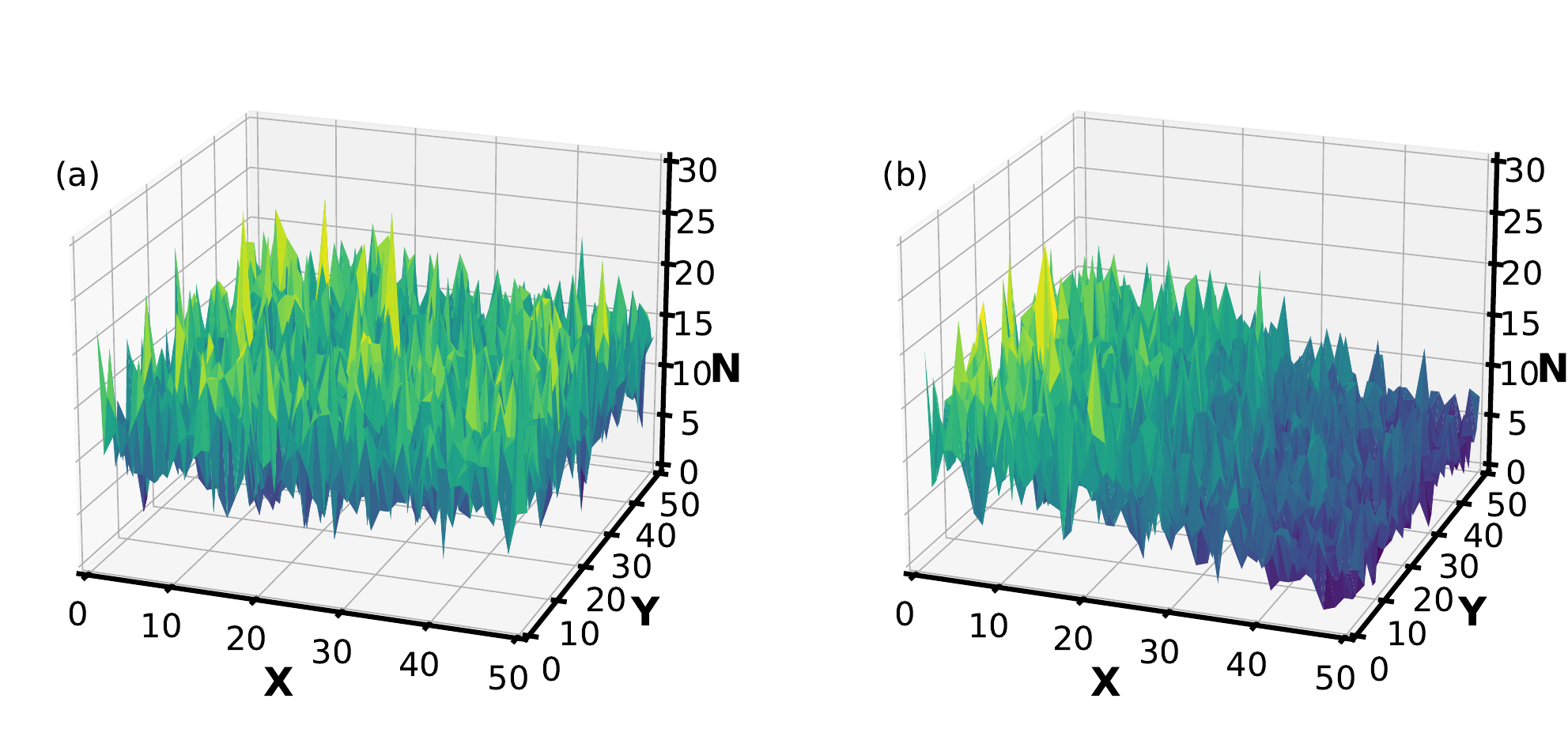}
    \caption{Number of trigger events for cells along an $x-y$ plane slice for which $n_z = 25$.  Left panel:  fast reconnection model without the sheared perturbation structure.  Right panel: fast reconnection model with the sheared perturbation structure. Color corresponds to number of events. }\label{fig:trigger} 
\end{center}
\end{figure*}

Finally, we assess the degree to which  the sheared structure of the pertubrations affects the triggering of reconnection events.  In Figure \ref{fig:trigger} we compare the number of triggered events that occurred in the $50\times 50$ cells along the $x-y$ plane slice for $n_z = 25$ for (a) the fast reconnection model presented in Section 5.1 and  (b) the first scenario considered in this section.  Statistical noise dominates the results due to the inherent randomness in the system. Nevertheless, no clear structure is evident in the trigger locations in the absence of a sheared injection. As evidenced by panel (b), however, sheared perturbation  produce a corresponding structure in the trigger profile.  Similar results are obtained for all other values of $n_z$, and the stacking of these planes decreases the signal to noise ratio for the obtained results.

\subsection{Slow Reconnection Model}

Motivated by the discussion at the end Section 5.1, our final simulation considers the extreme scenario for which $\Delta t_a = \Delta t_p$, so that a perturbation is added randomly in the lattice after every avalanche step.

Our results, shown in Figure \ref{fig:comparen}, indicate that the slopes of the output measure distributions are quite similar to those of the fast reconnection model.  However, due to the addition of perturbations at each avalanche time step, there are slightly more high-energy events in the slow reconnection model.
These differences are modest and only manifest themselves in the tails of the distributions. This result is expected, as the tails often converge more slowly than the rest of the distribution. Unfortunately, however, the observational verification of this difference would be difficult. The number of flares at the high end of the distribution is about six order of magnitude smaller than at low energies. 

On a related note, no current measurements exist for the distributions of flares from magnetized circumstellar disks. Our results thus provide a prediction: namely that the flare distribution should have a nearly universal and power-law form (as shown in the Figures). This result in not unexpected, as observations of solar flares show power-law distributions of energy, duration, and power \citep{Lu1991}. In addition, many astrophysical systems display flaring activity with power-law forms, including stellar flares \citep{balona2015}, soft gamma-ray repeaters \citep{cheng1996}, and gamma-ray bursts (\citealt{wang2013}; see also the review of \citealt{aschwanden2016}).

\bigskip 
$\,$
\bigskip 

\section{Summary and Discussion} \label{sec:results}

Motivated by the possibility of cosmic ray acceleration in the truncation region of circumstellar disks, this paper has generalized existing models of magnetic reconnection events to include Keplerian shear and additional complications. The main results of this work support three principle conclusions. First, we have shown that the magnetic truncation region can indeed support turbulent driven self-critical states that produce flaring events with power-law distributions of energy, duration, and power. Significantly, this ability to reach self-criticality is not compromised by the sheared magnetic field structure that may be imposed by the Keplerian rotation curve of the disk.

Second, the self-critical states that characterize the distributions of flaring events appear to be universal.  Indeed, the universality of the self-critical states found in earlier work is extended by the present analysis.  Previous work indicated a universality with respect to the perturbation parameters $\Delta b$ and $b_0$, the triggering value of the field $B_c$, and the number of cells in the lattice for $N\gtrsim50$ \citep{Lu1991,Lu1993,Edney1998}. In addition, our results show that the frequency distributions of avalanche energy that characterize flaring events are similar for: 1) both fast and slow reconnection processes; 2) periodic and external boundary conditions imposed on the magnetic field; and 3) the inclusion of a spatial structure in the perturbations and triggering conditions (where this latter generalization is motivated by the rotation curves of circumstellar disks). This universality seemingly justifies drawing parallels between well studied solar flare phenomena and the less well-understood disk flaring phenomena inferred in forming star/disk systems. 

Third, flaring models require $\sim 100$ G magnetic fields, similar to those found in solar flare environments, in order to generate the X-ray and cosmic ray luminosities inferred for young star/disk systems.  In addition, the reconnection time within a cell ($\sim 10$ s) indicates that the vast majority of avalanche events are shorter than the time scale on which perturbations are added to the lattice.  

As a final note, the smooth Laplace surfaces presented in this work provide a new method of exploring the self-critical problem (see Figure 2). These surfaces provide initial conditions that allow a system to be driven efficiently to its self-critical state.  However, we have also shown that these surfaces provide only a heuristic description of the true magnetic structure. 

The results of this work show that flaring activity produced in circumstellar disks -- operating in the disk truncation region -- proceed much like that occuring  on stellar surfaces. In particular, the presence of differential rotation and alternate boundary conditions do not prevent the system from reaching a state of self-organized criticality. This analogy between stellar and disk flaring activity has been implicitly assumed in many previous applications (e.g., see  \citealt{Shu1996,Shu2001,Gounelle2001,Leya2003,Duprat2007,adams2021}) and has been used to argue that circumstellar disks are likely to produce large fluxes of cosmic rays. This present work thus vindicates these previous assumptions.  

Nevertheless, this calculation represents only one step toward a full understanding of reconnection events in circumstellar disks, and a number of directions for future research are indicated. Here we have used a simple but physically motivated model \citep{Lu1991} to explore the effects of varying background conditions. Future work should study the problem with full MHD simulations  (although the reconnection problem in MHD remains difficult). This paper considers the energy, duration, and power of magnetic reconnection events and finds nearly universal power-law behavior. Another future direction would be to use these distributions of flaring properties to determine the corresponding distributions of cosmic rays that are accelerated through these events. In addition, once produced, the cosmic rays must propagate through both the local reconnection region of the disk and the larger-scale magnetic field structure of the young stellar object. Finally, with cosmic ray production and propagation determined, their effects on ionization of the disk and nuclear processing through spallation must be determined. 


\bigskip 
\centerline{Acknowledgements}
\medskip 

MF is grateful for support from the Hauck foundation. FCA is supported, in part, by the Leinweber Center for Theoretical Physics at the University of Michigan. ADF acknowledges support by the National Science Foundation Graduate Research Fellowship Program under Grant No. (DGE-1746045). DZS acknowledges financial support from the National Science Foundation  Grant No. AST-2107796, NASA Grant No. 80NSSC19K0444 and NASA Contract  NNX17AL71A from the NASA Goddard Spaceflight Center.

\bibliography{sample631}{}

\begin{thebibliography}{}
\expandafter\ifx\csname natexlab\endcsname\relax\def\natexlab#1{#1}\fi
\providecommand{\url}[1]{\href{#1}{#1}}
\providecommand{\dodoi}[1]{doi:~\href{http://doi.org/#1}{\nolinkurl{#1}}}
\providecommand{\doeprint}[1]{\href{http://ascl.net/#1}{\nolinkurl{http://ascl.net/#1}}}
\providecommand{\doarXiv}[1]{\href{https://arxiv.org/abs/#1}{\nolinkurl{https://arxiv.org/abs/#1}}}

\bibitem[{{Adams}(2010)}]{adams2010}
{Adams}, F.~C. 2010, \araa, 48, 47, \dodoi{10.1146/annurev-astro-081309-130830}

\bibitem[{{Adams}(2021)}]{adams2021}
---. 2021, \apj, 919, 10, \dodoi{10.3847/1538-4357/ac1111}

\bibitem[{{Akeson} {et~al.}(2005){Akeson}, {Walker}, {Wood}, {Eisner}, {Scire},
  {Penprase}, {Ciardi}, {van Belle}, {Whitney}, \& {Bjorkman}}]{Akeson2005}
{Akeson}, R.~L., {Walker}, C.~H., {Wood}, K., {et~al.} 2005, \apj, 622, 440,
  \dodoi{10.1086/427770}

\bibitem[{{Aschwanden} {et~al.}(1998){Aschwanden}, {Dennis}, \&
  {Benz}}]{Aschwanden1998}
{Aschwanden}, M.~J., {Dennis}, B.~R., \& {Benz}, A.~O. 1998, \apj, 497, 972,
  \dodoi{10.1086/305484}

\bibitem[{{Aschwanden} {et~al.}(2016){Aschwanden}, {Crosby}, {Dimitropoulou},
  {Georgoulis}, {Hergarten}, {McAteer}, {Milovanov}, {Mineshige}, {Morales},
  {Nishizuka}, {Pruessner}, {Sanchez}, {Sharma}, {Strugarek}, \&
  {Uritsky}}]{aschwanden2016}
{Aschwanden}, M.~J., {Crosby}, N.~B., {Dimitropoulou}, M., {et~al.} 2016, \ssr,
  198, 47, \dodoi{10.1007/s11214-014-0054-6}

\bibitem[{{Babcock} \& {Westervelt}(1990)}]{Babcock1990}
{Babcock}, K.~L., \& {Westervelt}, R.~M. 1990, \prl, 64, 2168,
  \dodoi{10.1103/PhysRevLett.64.2168}

\bibitem[{{Bak} {et~al.}(1987){Bak}, {Tang}, \& {Wiesenfeld}}]{bak1987}
{Bak}, P., {Tang}, C., \& {Wiesenfeld}, K. 1987, \prl, 59, 381,
  \dodoi{10.1103/PhysRevLett.59.381}

\bibitem[{{Bak} {et~al.}(1988){Bak}, {Tang}, \& {Wiesenfeld}}]{bak1988}
---. 1988, \pra, 38, 364, \dodoi{10.1103/PhysRevA.38.364}

\bibitem[{{Balona}(2015)}]{balona2015}
{Balona}, L.~A. 2015, \mnras, 447, 2714, \dodoi{10.1093/mnras/stu2651}

\bibitem[{{Bate}(2009)}]{Bate2009}
{Bate}, M.~R. 2009, \mnras, 392, 590, \dodoi{10.1111/j.1365-2966.2008.14106.x}

\bibitem[{{Cameron} \& {Truran}(1977)}]{Cameron1977}
{Cameron}, A.~G.~W., \& {Truran}, J.~W. 1977, \icarus, 30, 447,
  \dodoi{10.1016/0019-1035(77)90101-4}

\bibitem[{{Charbonneau} {et~al.}(2001){Charbonneau}, {McIntosh}, {Liu}, \&
  {Bogdan}}]{Charbonneau2001}
{Charbonneau}, P., {McIntosh}, S.~W., {Liu}, H.-L., \& {Bogdan}, T.~J. 2001,
  \solphys, 203, 321, \dodoi{10.1023/A:1013301521745}

\bibitem[{{Cheng} {et~al.}(1996){Cheng}, {Epstein}, {Guyer}, \&
  {Young}}]{cheng1996}
{Cheng}, B., {Epstein}, R.~I., {Guyer}, R.~A., \& {Young}, A.~C. 1996, \nat,
  382, 518, \dodoi{10.1038/382518a0}

\bibitem[{{Crosby} {et~al.}(1993){Crosby}, {Aschwanden}, \&
  {Dennis}}]{Crosby1993}
{Crosby}, N.~B., {Aschwanden}, M.~J., \& {Dennis}, B.~R. 1993, \solphys, 143,
  275, \dodoi{10.1007/BF00646488}

\bibitem[{{D'Alessio} {et~al.}(2005){D'Alessio}, {Hartmann}, {Calvet},
  {Franco-Hern{\'a}ndez}, {Forrest}, {Sargent}, {Furlan}, {Uchida}, {Green},
  {Watson}, {Chen}, {Kemper}, {Sloan}, \& {Najita}}]{DAlessio2005}
{D'Alessio}, P., {Hartmann}, L., {Calvet}, N., {et~al.} 2005, \apj, 621, 461,
  \dodoi{10.1086/427490}

\bibitem[{{Dawson} \& {Johnson}(2018)}]{dawson2018}
{Dawson}, R.~I., \& {Johnson}, J.~A. 2018, \araa, 56, 175,
  \dodoi{10.1146/annurev-astro-081817-051853}

\bibitem[{{de Arcangelis} {et~al.}(2006){de Arcangelis}, {Godano}, {Lippiello},
  \& {Nicodemi}}]{deArcangelis06}
{de Arcangelis}, L., {Godano}, C., {Lippiello}, E., \& {Nicodemi}, M. 2006,
  \prl, 96, 051102, \dodoi{10.1103/PhysRevLett.96.051102}

\bibitem[{{Dmitruk} \& {G{\'o}mez}(1997)}]{Dmitruk1997}
{Dmitruk}, P., \& {G{\'o}mez}, D.~O. 1997, \apjl, 484, L83,
  \dodoi{10.1086/310760}

\bibitem[{{Duprat} \& {Tatischeff}(2007)}]{Duprat2007}
{Duprat}, J., \& {Tatischeff}, V. 2007, \apjl, 671, L69, \dodoi{10.1086/524297}

\bibitem[{Edney {et~al.}(1998)Edney, Robinson, \& Chisholm}]{Edney1998}
Edney, S.~D., Robinson, P.~A., \& Chisholm, D. 1998, Phys. Rev. E, 58, 5395,
  \dodoi{10.1103/PhysRevE.58.5395}

\bibitem[{{Einaudi} \& {Velli}(1999)}]{Einaudi1999}
{Einaudi}, G., \& {Velli}, M. 1999, Physics of Plasmas, 6, 4146,
  \dodoi{10.1063/1.873679}

\bibitem[{{Einaudi} {et~al.}(1996){Einaudi}, {Velli}, {Politano}, \&
  {Pouquet}}]{Einaudi1996}
{Einaudi}, G., {Velli}, M., {Politano}, H., \& {Pouquet}, A. 1996, \apjl, 457,
  L113, \dodoi{10.1086/309893}

\bibitem[{Farhang {et~al.}(2018)Farhang, Safari, \& Wheatland}]{Farhang_2018}
Farhang, N., Safari, H., \& Wheatland, M.~S. 2018, The Astrophysical Journal,
  859, 41, \dodoi{10.3847/1538-4357/aac01b}

\bibitem[{{Feigelson} {et~al.}(2005){Feigelson}, {Getman}, {Townsley},
  {Garmire}, {Preibisch}, {Grosso}, {Montmerle}, {Muench}, \&
  {McCaughrean}}]{feigelson2005}
{Feigelson}, E.~D., {Getman}, K., {Townsley}, L., {et~al.} 2005, \apjs, 160,
  379, \dodoi{10.1086/432512}

\bibitem[{{Feinstein} {et~al.}(2022){Feinstein}, {Seligman}, {G{\"u}nther}, \&
  {Adams}}]{Feinstein2022}
{Feinstein}, A.~D., {Seligman}, D.~Z., {G{\"u}nther}, M.~N., \& {Adams}, F.~C.
  2022, \apjl, 925, L9, \dodoi{10.3847/2041-8213/ac4b5e}

\bibitem[{{Fujii} \& {Kimura}(2022)}]{fujii2022}
{Fujii}, Y.~I., \& {Kimura}, S.~S. 2022, \apjl, 937, L37,
  \dodoi{10.3847/2041-8213/ac86c2}

\bibitem[{{Gaches} {et~al.}(2020){Gaches}, {Walch}, {Offner}, \&
  {M{\"u}nker}}]{gaches2020}
{Gaches}, B. A.~L., {Walch}, S., {Offner}, S. S.~R., \& {M{\"u}nker}, C. 2020,
  \apj, 898, 79, \dodoi{10.3847/1538-4357/ab9a38}

\bibitem[{{Galsgaard} \& {Nordlund}(1996)}]{Galsgaard1996}
{Galsgaard}, K., \& {Nordlund}, {\r{A}}. 1996, \jgr, 101, 13445,
  \dodoi{10.1029/96JA00428}

\bibitem[{{Galtier}(1999)}]{Galtier1999}
{Galtier}, S. 1999, \apj, 521, 483, \dodoi{10.1086/307537}

\bibitem[{{Galtier} \& {Pouquet}(1998)}]{Galtier1998}
{Galtier}, S., \& {Pouquet}, A. 1998, \solphys, 179, 141,
  \dodoi{10.1023/A:1005056102064}

\bibitem[{{Georgoulis} {et~al.}(1998){Georgoulis}, {Velli}, \&
  {Einaudi}}]{Georgoulis1998}
{Georgoulis}, M.~K., {Velli}, M., \& {Einaudi}, G. 1998, \apj, 497, 957,
  \dodoi{10.1086/305486}

\bibitem[{{Goswami} {et~al.}(2005){Goswami}, {Marhas}, {Chaussidon},
  {Gounelle}, \& {Meyer}}]{Goswami2005}
{Goswami}, J.~N., {Marhas}, K.~K., {Chaussidon}, M., {Gounelle}, M., \&
  {Meyer}, B.~S. 2005, in Astronomical Society of the Pacific Conference
  Series, Vol. 341, Chondrites and the Protoplanetary Disk, ed. A.~N. {Krot},
  E.~R.~D. {Scott}, \& B.~{Reipurth}, 485

\bibitem[{{Gounelle} {et~al.}(2001){Gounelle}, {Shu}, {Shang}, {Glassgold},
  {Rehm}, \& {Lee}}]{Gounelle2001}
{Gounelle}, M., {Shu}, F.~H., {Shang}, H., {et~al.} 2001, \apj, 548, 1051,
  \dodoi{10.1086/319019}

\bibitem[{{Grimm} \& {McSween}(1993)}]{Grimm1993}
{Grimm}, R.~E., \& {McSween}, H.~Y. 1993, Science, 259, 653

\bibitem[{{Hayashi} \& {Nakano}(1965)}]{Hayashi1965}
{Hayashi}, C., \& {Nakano}, T. 1965, Progress of Theoretical Physics, 34, 754,
  \dodoi{10.1143/PTP.34.754}

\bibitem[{{Hester} {et~al.}(2004){Hester}, {Desch}, {Healy}, \&
  {Leshin}}]{Hester2004}
{Hester}, J.~J., {Desch}, S.~J., {Healy}, K.~R., \& {Leshin}, L.~A. 2004,
  Science, 304, 1116, \dodoi{10.1126/science.1096808}

\bibitem[{{Hevey} \& {Sanders}(2006)}]{Hevey2006}
{Hevey}, P.~J., \& {Sanders}, I.~S. 2006, \maps, 41, 95,
  \dodoi{10.1111/j.1945-5100.2006.tb00195.x}

\bibitem[{{Ikoma} {et~al.}(2018){Ikoma}, {Elkins-Tanton}, {Hamano}, \&
  {Suckale}}]{Ikoma2018}
{Ikoma}, M., {Elkins-Tanton}, L., {Hamano}, K., \& {Suckale}, J. 2018, \ssr,
  214, 76, \dodoi{10.1007/s11214-018-0508-3}

\bibitem[{{Ionson}(1985)}]{Ionson1985}
{Ionson}, J.~A. 1985, \solphys, 100, 289, \dodoi{10.1007/BF00158433}

\bibitem[{{Johns-Krull}(2007)}]{Johns2007}
{Johns-Krull}, C.~M. 2007, \apj, 664, 975, \dodoi{10.1086/519017}

\bibitem[{{Johns-Krull} {et~al.}(2009){Johns-Krull}, {Greene}, {Doppmann}, \&
  {Covey}}]{Johns-Krull2009}
{Johns-Krull}, C.~M., {Greene}, T.~P., {Doppmann}, G.~W., \& {Covey}, K.~R.
  2009, \apj, 700, 1440, \dodoi{10.1088/0004-637X/700/2/1440}

\bibitem[{{Kadanoff} {et~al.}(1989){Kadanoff}, {Nagel}, {Wu}, \&
  {Zhou}}]{Kadanoff1989}
{Kadanoff}, L.~P., {Nagel}, S.~R., {Wu}, L., \& {Zhou}, S.-M. 1989, \pra, 39,
  6524, \dodoi{10.1103/PhysRevA.39.6524}

\bibitem[{{Krot} {et~al.}(2009){Krot}, {Amelin}, {Bland}, {Ciesla}, {Connelly},
  {Davis}, {Huss}, {Hutcheon}, {Makide}, {Nagashima}, {Nyquist}, {Russell},
  {Scott}, {Thrane}, {Yurimoto}, \& {Yin}}]{Krot2009}
{Krot}, A.~N., {Amelin}, Y., {Bland}, P., {et~al.} 2009, \gca, 73, 4963,
  \dodoi{10.1016/j.gca.2008.09.039}

\bibitem[{{LaTourrette} \& {Wasserburg}(1998)}]{LaTourrette1998}
{LaTourrette}, T., \& {Wasserburg}, G.~J. 1998, Earth and Planetary Science
  Letters, 158, 91, \dodoi{10.1016/S0012-821X(98)00048-X}

\bibitem[{{Lee} {et~al.}(1998){Lee}, {Shu}, {Shang}, {Glassgold}, \&
  {Rehm}}]{Lee1998}
{Lee}, T., {Shu}, F.~H., {Shang}, H., {Glassgold}, A.~E., \& {Rehm}, K.~E.
  1998, \apj, 506, 898, \dodoi{10.1086/306284}

\bibitem[{{Leya} {et~al.}(2003){Leya}, {Halliday}, \& {Wieler}}]{Leya2003}
{Leya}, I., {Halliday}, A.~N., \& {Wieler}, R. 2003, \apj, 594, 605,
  \dodoi{10.1086/376795}

\bibitem[{{Lichtenberg} {et~al.}(2019){Lichtenberg}, {Golabek}, {Burn},
  {Meyer}, {Alibert}, {Gerya}, \& {Mordasini}}]{Lichtenberg2019}
{Lichtenberg}, T., {Golabek}, G.~J., {Burn}, R., {et~al.} 2019, Nature
  Astronomy, 3, 307, \dodoi{10.1038/s41550-018-0688-5}

\bibitem[{{Litvinenko}(1996)}]{Litvinenko1996}
{Litvinenko}, Y.~E. 1996, \solphys, 167, 321, \dodoi{10.1007/BF00146342}

\bibitem[{{Longcope} \& {Sudan}(1994)}]{Longcope1994}
{Longcope}, D.~W., \& {Sudan}, R.~N. 1994, \apj, 437, 491,
  \dodoi{10.1086/175013}

\bibitem[{{Lu} \& {Hamilton}(1991)}]{Lu1991}
{Lu}, E.~T., \& {Hamilton}, R.~J. 1991, \apjl, 380, L89, \dodoi{10.1086/186180}

\bibitem[{{Lu} {et~al.}(1993){Lu}, {Hamilton}, {McTiernan}, \&
  {Bromund}}]{Lu1993}
{Lu}, E.~T., {Hamilton}, R.~J., {McTiernan}, J.~M., \& {Bromund}, K.~R. 1993,
  \apj, 412, 841, \dodoi{10.1086/172966}

\bibitem[{{McKee} \& {Ostriker}(2007)}]{mckeeost}
{McKee}, C.~F., \& {Ostriker}, E.~C. 2007, \araa, 45, 565,
  \dodoi{10.1146/annurev.astro.45.051806.110602}

\bibitem[{{Morales} \& {Santos}(2020)}]{Morales2020}
{Morales}, L.~F., \& {Santos}, N.~A. 2020, \solphys, 295, 155,
  \dodoi{10.1007/s11207-020-01713-0}

\bibitem[{{Moskovitz} \& {Gaidos}(2011)}]{Moskovitz2011}
{Moskovitz}, N., \& {Gaidos}, E. 2011, \maps, 46, 903,
  \dodoi{10.1111/j.1945-5100.2011.01201.x}

\bibitem[{{Newman}(2005)}]{newmann2005}
{Newman}, M.~E.~J. 2005, Contemporary Physics, 46, 323,
  \dodoi{10.1080/00107510500052444}

\bibitem[{{Newman} \& {Sneppen}(1996)}]{Newman1996}
{Newman}, M.~E.~J., \& {Sneppen}, K. 1996, \pre, 54, 6226,
  \dodoi{10.1103/PhysRevE.54.6226}

\bibitem[{{Norman} {et~al.}(2001){Norman}, {Charbonneau}, {McIntosh}, \&
  {Liu}}]{norman2001}
{Norman}, J.~P., {Charbonneau}, P., {McIntosh}, S.~W., \& {Liu}, H.-L. 2001,
  \apj, 557, 891, \dodoi{10.1086/321678}

\bibitem[{{Padovani} {et~al.}(2016){Padovani}, {Marcowith}, {Hennebelle}, \&
  {Ferri{\`e}re}}]{padovani2016}
{Padovani}, M., {Marcowith}, A., {Hennebelle}, P., \& {Ferri{\`e}re}, K. 2016,
  \aap, 590, A8, \dodoi{10.1051/0004-6361/201628221}

\bibitem[{Parker(1979)}]{Parker1979}
Parker, E. 1979, {Cosmical magnetic fields. Their origin and their activity
  (Oxford: Oxford Univ. Press}

\bibitem[{{Parker}(1970)}]{Parker1970}
{Parker}, E.~N. 1970, \araa, 8, 1, \dodoi{10.1146/annurev.aa.08.090170.000245}

\bibitem[{{Parker}(1972)}]{Parker1972}
---. 1972, \apj, 174, 499, \dodoi{10.1086/151512}

\bibitem[{{Patterson}(1994)}]{Patterson1994}
{Patterson}, J. 1994, \pasp, 106, 209, \dodoi{10.1086/133375}

\bibitem[{{Preibisch} {et~al.}(2005){Preibisch}, {Kim}, {Favata}, {Feigelson},
  {Flaccomio}, {Getman}, {Micela}, {Sciortino}, {Stassun}, {Stelzer}, \&
  {Zinnecker}}]{preibisch2005}
{Preibisch}, T., {Kim}, Y.-C., {Favata}, F., {et~al.} 2005, \apjs, 160, 401,
  \dodoi{10.1086/432891}

\bibitem[{{Reiter}(2020)}]{Reiter2020}
{Reiter}, M. 2020, \aap, 644, L1, \dodoi{10.1051/0004-6361/202039334}

\bibitem[{{Rosner} \& {Vaiana}(1978)}]{Rosner1978}
{Rosner}, R., \& {Vaiana}, G.~S. 1978, \apj, 222, 1104, \dodoi{10.1086/156227}

\bibitem[{{Schramm}(1971)}]{Schramm1971}
{Schramm}, D.~N. 1971, \apss, 13, 249, \dodoi{10.1007/BF00656331}

\bibitem[{{Shu} {et~al.}(1994){Shu}, {Najita}, {Ostriker}, {Wilkin}, {Ruden},
  \& {Lizano}}]{Shu1994}
{Shu}, F., {Najita}, J., {Ostriker}, E., {et~al.} 1994, \apj, 429, 781,
  \dodoi{10.1086/174363}

\bibitem[{{Shu} {et~al.}(1987){Shu}, {Adams}, \& {Lizano}}]{Shu1987}
{Shu}, F.~H., {Adams}, F.~C., \& {Lizano}, S. 1987, \araa, 25, 23,
  \dodoi{10.1146/annurev.aa.25.090187.000323}

\bibitem[{{Shu} {et~al.}(2001){Shu}, {Shang}, {Gounelle}, {Glassgold}, \&
  {Lee}}]{Shu2001}
{Shu}, F.~H., {Shang}, H., {Gounelle}, M., {Glassgold}, A.~E., \& {Lee}, T.
  2001, \apj, 548, 1029, \dodoi{10.1086/319018}

\bibitem[{{Shu} {et~al.}(1996){Shu}, {Shang}, \& {Lee}}]{Shu1996}
{Shu}, F.~H., {Shang}, H., \& {Lee}, T. 1996, Science, 271, 1545,
  \dodoi{10.1126/science.271.5255.1545}

\bibitem[{{Urey}(1955)}]{Urey1955}
{Urey}, H.~C. 1955, Proceedings of the National Academy of Science, 41, 127,
  \dodoi{10.1073/pnas.41.3.127}

\bibitem[{{Wang} \& {Dai}(2013)}]{wang2013}
{Wang}, F.~Y., \& {Dai}, Z.~G. 2013, Nature Physics, 9, 465,
  \dodoi{10.1038/nphys2670}

\bibitem[{{Weiss} {et~al.}(2022){Weiss}, {Millholland}, {Petigura}, {Adams},
  {Batygin}, {Bloch}, \& {Mordasini}}]{weiss2022}
{Weiss}, L.~M., {Millholland}, S.~C., {Petigura}, E.~A., {et~al.} 2022, arXiv
  e-prints, arXiv:2203.10076, \dodoi{10.48550/arXiv.2203.10076}

\bibitem[{{Weiss} {et~al.}(2018){Weiss}, {Marcy}, {Petigura}, {Fulton},
  {Howard}, {Winn}, {Isaacson}, {Morton}, {Hirsch}, {Sinukoff}, {Cumming},
  {Hebb}, \& {Cargile}}]{weiss2018}
{Weiss}, L.~M., {Marcy}, G.~W., {Petigura}, E.~A., {et~al.} 2018, \aj, 155, 48,
  \dodoi{10.3847/1538-3881/aa9ff6}

\end{thebibliography}
\bibliographystyle{aasjournal}

\end{document}